\documentclass[amsthm,a4]{PoS}

\usepackage{bm}
\usepackage{graphicx}
\usepackage{amsmath}
\usepackage{amsthm}
\usepackage{amsfonts}
\usepackage{amssymb}
\newcommand{\tht}{\textheight}
\newcommand{\ig}{\includegraphics}

\title{Multi-hadron Operators with All-to-All Quark  Propagators}

\ShortTitle{Multi-hadron propagators}

\author{
  \hspace*{5mm}J.~Bulava,$^a$, 
  R.G.~Edwards,$^b$ 
  \speaker{K.J.~Juge},$^c$ 
  C.J.~Morningstar,$^a$
  M.J.~Peardon $^d$ \\
  \centerline{(for the Hadron Spectrum Collaboration)}\\

\hspace*{8mm}\llap{$^a$}Department of Physics, Carnegie Mellon University, Pittsburgh, PA 15213, USA\\
\hspace*{8mm}\llap{$^b$}Thomas Jefferson National Accelerator Facility, Newport News, VA 23606, USA\\
\hspace*{8mm}\llap{$^c$}Department of Physics, University of the Pacific, Stockton, CA 95211, USA\\
\hspace*{8mm}\llap{$^d$}School of Mathematics, Trinity College, Dublin, Ireland

}
\abstract{Hadron spectroscopy on dynamical configurations are faced with the difficulties of dealing with the mixing of single particle states and multi-hadron states (for large spatial volumes and light dynamical quarks masses). It is conceivable that explicit multi-hadron interpolating operators will be necessary for obtaining sufficiently good overlap on to multi-hadron states in order to extract the low-lying excitation spectrum. We explore here the feasibility of using four noise diluted all-to-all quark propagators in the construction of explicit two-hadron operators on quenched, anisotropic lattices. Our longer term goal is to use these operators on large anisotropic, dynamical configurations for hadron spectroscopy.
}

\FullConference{The XXVI International Symposium on Lattice Field Theory\\
		            July 14-19 August 2008\\
		            Williamsburg, Virginia, USA}

\begin{document}

\section{Introduction}

The goal of the Hadron Spectrum Collaboration is to determine the low-lying hadron spectrum from first principles lattice QCD simulations. The low-lying spectra of baryons in the quenched approximation has been reported in Refs.~\cite{Zanotti:2003fx}. There has been progress made in dynamical simulations in the last few years by many collaborations as more and more light, dynamical confurations had become available (See \cite{McNeile:2007fu} for recent reviews). 

One of the fundamental challenges in spectroscopy of excited states is the issue of identifying multi-particle states in the spectrum. It is our expectation that explicit multi-particle operators will become necessary to determine the spectrum above multi-particle thresholds on (large) light, dynamical configurations. The shift in energy in a finite volume of two-particle states \cite{Luscher:1986pf} may suggest that finite momentum operators may have better overlap when the interaction between the two particles is not negligible. Simulation of correlation matrices with finite momenta operators using point-to-all quark propagators do not appear to be more cost effective than using stochastic estimates for all-to-all propagators with some improvement scheme. Moreover, the number of contractions increase dramatically for two-particle correlation functions constructed from point-to-all propagators which may result in a significant fraction of the simulation time being used for analysis. For these reasons, it is favourable to use some stochastic estimate of all-to-all quark propagators (\cite{Bitar:1988bb}-\cite{Hashimoto:2008xg}) in the simulation of low-lying multi-particle spectra.

In this study, we continue our efforts on examing the efficacy of the ``dilution method" (\cite{Bernardson:1993yg,Wilcox:1999ab,Foley:2005ac}) for estimating all-to-all quark propagators for the purpose of simulating multi-hadron states. Earlier works have demonstrated the efficacy of this method for zero momentum meson operators and group-theoretically motivated baryon operators \cite{Edwards:2007en}. A detailed study of the dilution scheme dependence of three-quark correlators has been reported in these proceedings \cite{john}. We shall focus here on finite momentum pion (two-quark) operators and $\pi^+\pi^+$ (four-quark) operators. 

\section{Construction of Operators/Correlators}

\subsection{Parameters}

We have used 100 quenched, anisotropic ($a_s/a_t=3$) Wilson gauge lattices ($12^3\times48$) at $\beta=6.1$ and quark masses which correspond to pion masses of approximately 700 MeV. The gauge fields were stout-smeared \cite{Morningstar:2003gk} with $(n,\epsilon)=(16,0.15625)$ and the quark fields were further smeared using gauge invariant Gaussian smearing with $(n,\sigma)=(32,3.0)$.

\subsection{Two-Quark Operators}

\begin{figure}[t]
\begin{center}
\begin{tabular}{cc}
\begin{minipage}{73mm}
\begin{center}
\ig[height=0.3\tht]{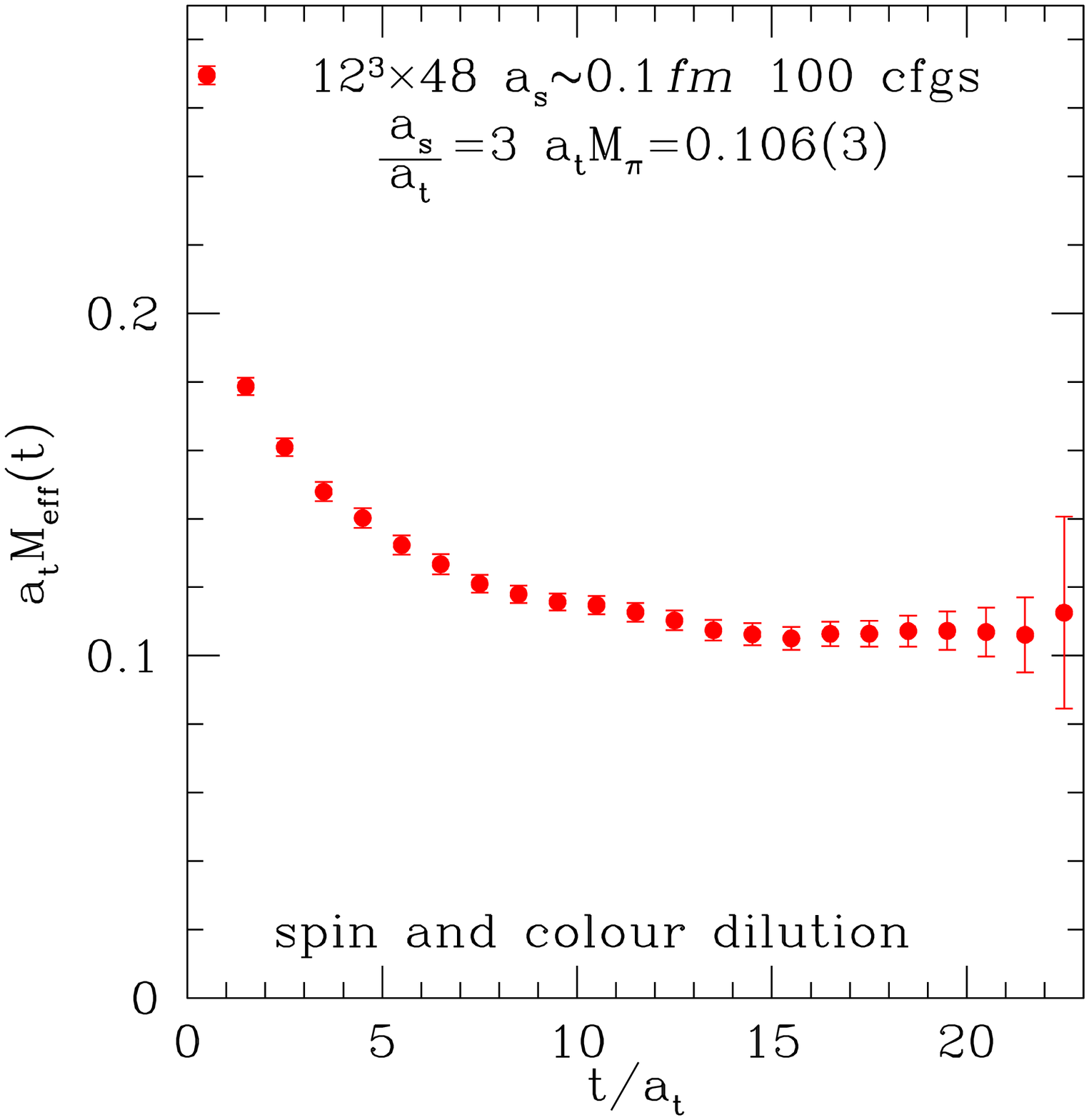} 
\caption{The pion effective mass using time-spin-colour diluted quark propagators. A single time-slice was used for the source operator and a periodic definition of the effective mass was used with $\Delta t=3$.}
\label{fig:pieffmass}
\end{center}
\end{minipage}&
\begin{minipage}{73mm}
\begin{center}
\ig[height=0.3\tht]{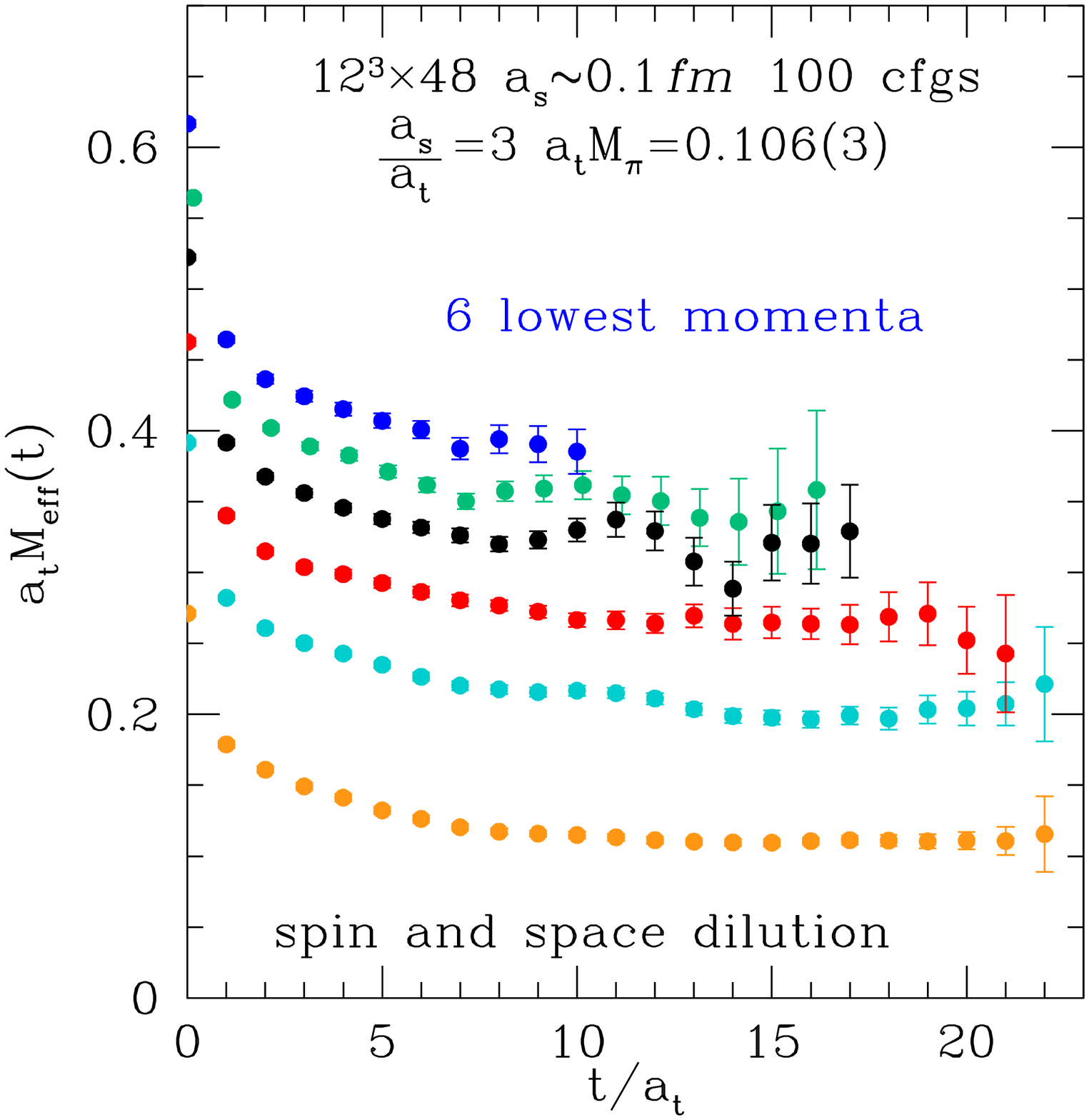} 
\caption{The pion effective masses for the first 6 lowest momenta with time-spin-space-diluted quark propagators. A single time-slice was used for the source operator and a periodic definition of the effective mass was used with $\Delta t=3$.} 
\end{center}
\label{fig:finiteP}
\end{minipage}
\end{tabular}
\end{center}
\end{figure}

Two independent $Z_4$ noise sources $\eta_{[A]}(\vec{x},t)$ and their solutions $\phi_{[A]}(\vec{x},t)$ (with $A=0,1$) were generated in order to construct the meson correlation functions. A meson correlation function is given by,
$$
C(\vec{p},t)=\sum_{t_0,\vec{x}_0,\vec{x}}\left[e^{-i\vec{p}\cdot(\vec{x}-\vec{x}_0)} Q(\vec{x},t+t_0;\vec{x_0},t_0)\Gamma^\dagger Q(\vec{x_0},t_0;\vec{x},t+t_0)\Gamma\right]
$$
where $Q(\vec{x},t+t_0;\vec{x_0},t_0)$ is the quark propagator from $(\vec{x_0},t_0)$ to $(\vec{x},t+t_0)$. In terms of the noise sources and solutions, this correlation function becomes
$$
C(\vec{p},t)=\sum_{t_0,\vec{x}_0,\vec{x}}\left[e^{-i\vec{p}\cdot(\vec{x}-\vec{x}_0)}\left(\eta^{\dagger}_{[0]}\Gamma\phi_{[1]}\right)(\vec{x},t)\left(\eta^{\dagger}_{[1]}\Gamma^\dagger\phi_{[0]}\right)(\vec{x}_0,t_0)\right]
$$
where the meson creation and annihilation operators (in parentheses) are contracted over the colour indices. This construction requires $N_t$ Dirac matrix inversions for each noise index $A$ since $\eta_{[A]}(\vec{x},t)$ must be given on all time-slices (and their corresponding solutions). This would have had the benefit of increased statistics as one has $N_t$ measurements on each lattice, but can be computationally demanding because the number of inversions increases by a factor of 48 (for this lattice). However, one can settle with the source on a single time-slice only, $t_0$, if we use $\gamma_5$ Hermiticity for the anti-quark propagator. In other words, we can write 
$$
C(\vec{p},t)=\sum_{t_0,\vec{x}_0,\vec{x}}\left[e^{-i\vec{p}\cdot(\vec{x}-\vec{x}_0)}\left(\phi^{\dagger}_{[0]}\gamma_5\Gamma\phi_{[1]}\right)(\vec{x},t)\left(\eta^{\dagger}_{[1]}(\gamma_5\Gamma)^\dagger\eta_{[0]}\right)(\vec{x}_0,t_0)\right]
$$
for the meson correlation function, which only contains the random noise source at $t_0$. The pion effective mass using time-spin-colour dilution with this method is shown in Fig.~\ref{fig:pieffmass}. The cost of producing such an effective mass is twice as much as that for traditional point-to-all propagators since two independent noise sources must be used in the stochastic case. 

It is just as simple to form correlation functions with finite momenta as with zero momentum since the correlation function is a product of meson creation and annihilation operators which can have any of the allowed momenta values. We construct the finite momenta operators given by,
\begin{eqnarray}\nonumber
\tilde{M}_{[0,1]}^{i,j}(\vec{p},t_0)&=&\sum_{\vec{x}}e^{i\vec{p}\cdot\vec{x}}\eta_{[0]}^\dagger\left(\gamma_5\Gamma\right)^\dagger\eta_{[1]}(\vec{x},t_0)\\\nonumber
M_{[1,0]}^{j,i}(\vec{p},t)&=&\sum_{\vec{x}}e^{-i\vec{p}\cdot\vec{x}}\phi_{[1]}^\dagger\gamma_5\Gamma\phi_{[0]}(\vec{x},t)
\end{eqnarray}
from which we make the finite $\vec{p}$ correlation functions (post-hoc). The effective pion mass for the six lowest momenta are shown in Fig.~\ref{fig:finiteP} where spin+space-cubic dilution was used for the stochastic estimates of the quark propagators. It is clear that one can measure up to (at least) five distinct momenta with 100 configurations. Note that the cost comparison with the point-to-all method is difficult here as this depends on how many source vectors were chosen for the creation operator in the point-to-all simulation. 

\begin{figure}[t]
\begin{center}
\begin{tabular}{cc}
\begin{minipage}{73mm}
\begin{center}
\ig[height=0.3\tht]{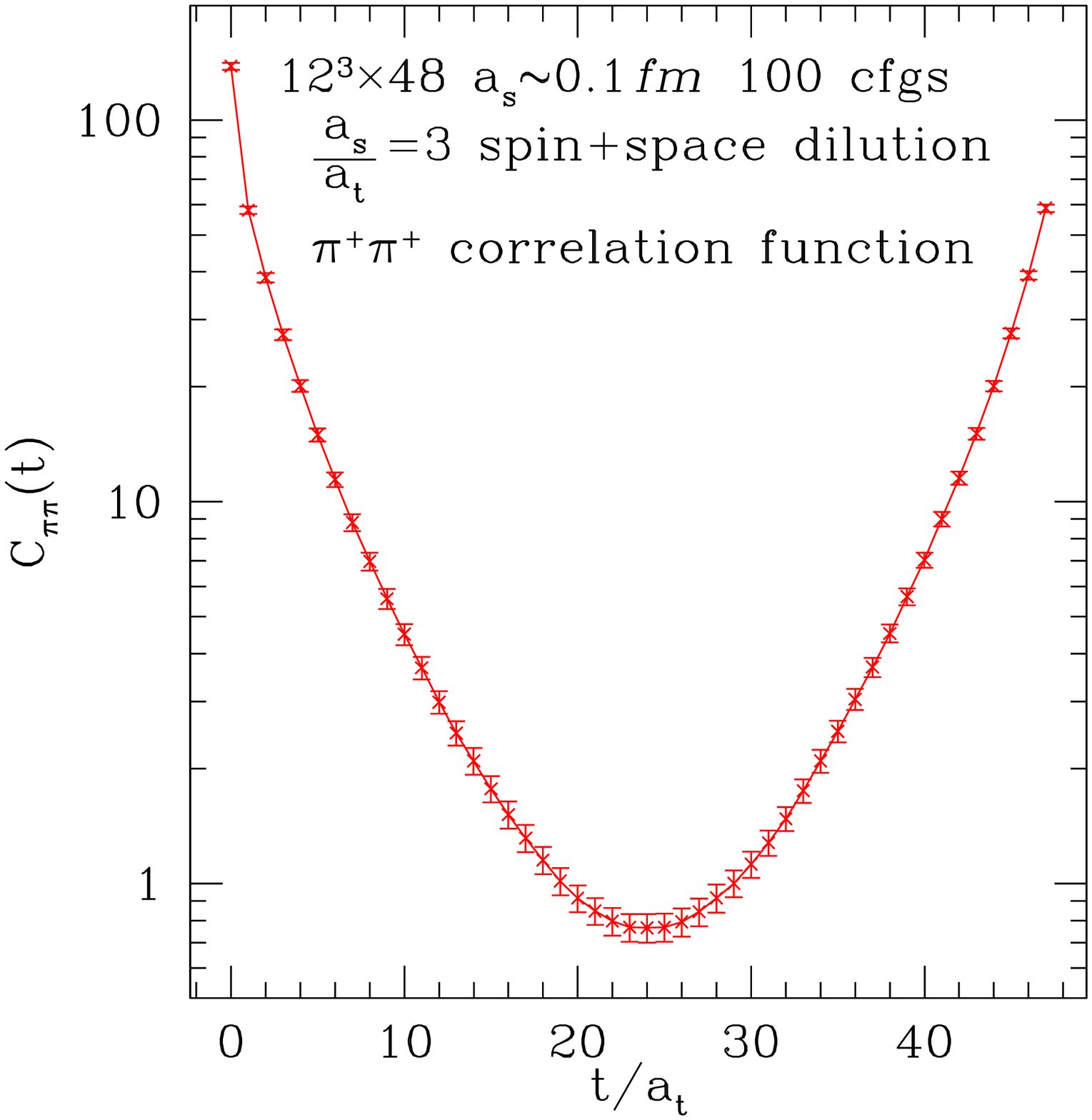}
\caption{The $\pi^+\pi^+$ correlation function with  time-spin-space-dilution.} 
\label{fig:pipicorr}
\end{center}
\end{minipage}&
\begin{minipage}{73mm}
\begin{center}
\ig[height=0.3\tht]{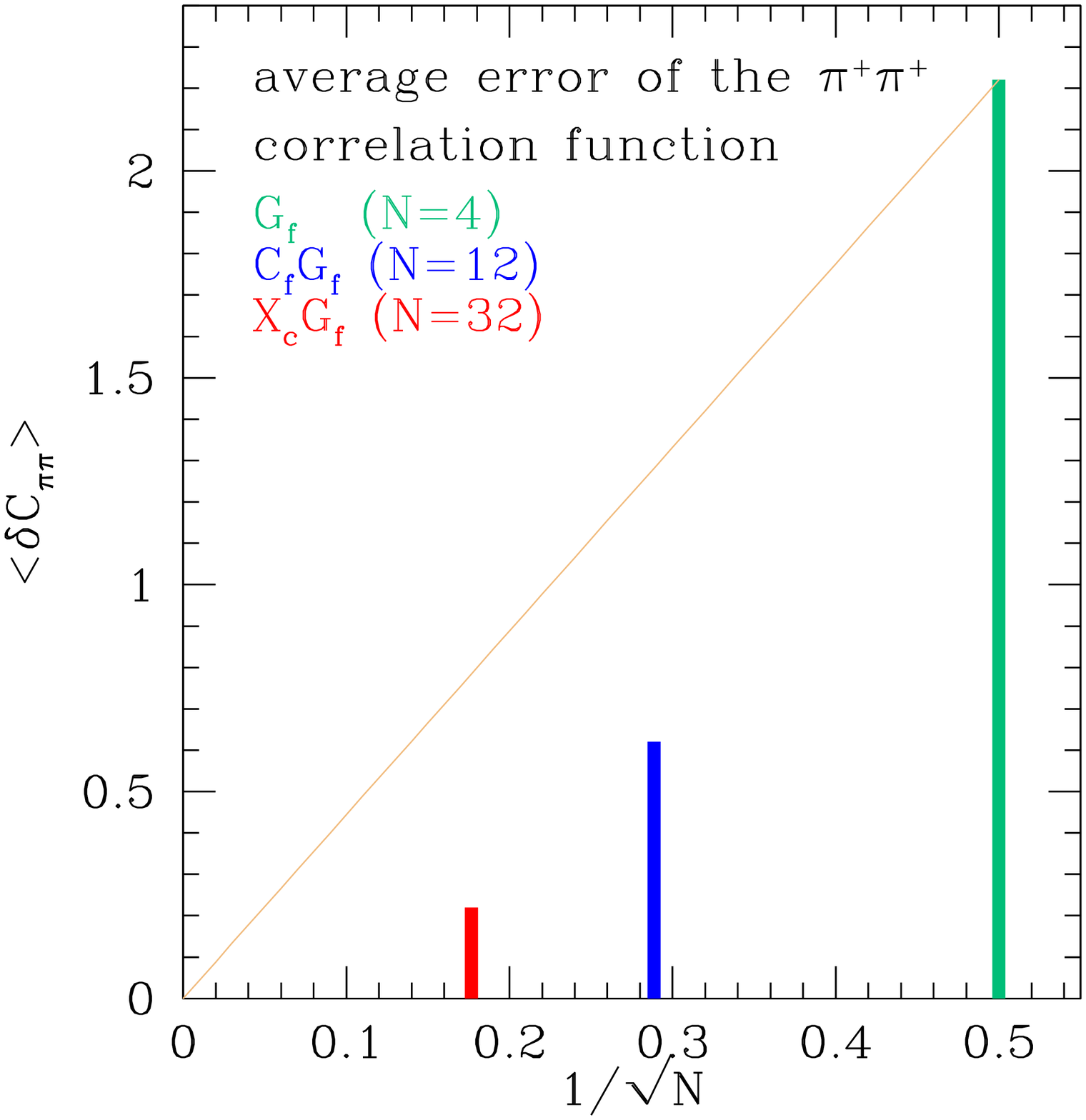} 
\caption{The average error of the $\pi^+\pi^+$ correlation function from timeslice 5 to 19 for various dilution schemes.} 
\label{fig:D-C}
\end{center}
\end{minipage}
\end{tabular}
\end{center}
\end{figure}

\subsection{Three-Quark Operators}

The dilution test for three-quark operators have been reported in \cite{Edwards:2007en} and \cite{john}. Diagonalization with several different nucleon-type operators was found to be stable at relatively low dilution levels. 

\subsection{Four-Quark Operators}

We have computed the $I=2$, $\pi\pi$ correlation function as an example of an operator involving four-quarks. The correlation function for this state has the form,
$$C_{\pi\pi}(t,t_0)=C^{(\pi)}_{[0,1]}C^{(\pi)}_{[2,3]}-C_{[0,1,2,3]}$$
where $C^{(\pi)}_{[A,B]}(t,t_0)$ is the pion correlation function constructed from noise sources $A$ and $B$ and the ``crossed" diagram is given by
$$C_{[0,1,2,3]}(t_0+t,t_0) = \tilde{M}_{[0,1]}^{i,j}(t_0)M_{[1,2]}^{j,k}(t_0+t)\tilde{M}_{[2,3]}^{k,l}(t_0)M_{[3,0]}^{l,i}(t_0+t)$$
We have used `zero-momentum' pion operators in this study, but any of the other finite momenta operators could have been used as well. (Note that the phase shift could also had been determined by diagonalizing the matrix of correlation functions using these `finite momenta' operators.) An example of the four-quark correlation function with spin and space-cubic dilution is shown in Fig.~\ref{fig:pipicorr}. 

\section{Simulation/Results}
\subsection{Parameters}
For this exploratory study, we have used the same quenched (100) configurations that were used in the excited baryon spectrum study \cite{Basak:2007kj}. These are the quenched, anisotropic ($a_s/a_t=3$) Wilson action lattices at $\beta=6.1, 12^3\times48$. The spatial lattice spacing is $~0.1$ fm. Anisotropic Wilson fermions with pions masses of roughly $700\ MeV$ were used for the valence quarks. 

\subsection{Two-Quark Operators}
\begin{figure}[t]
\begin{center}
\begin{tabular}{cc}
\begin{minipage}{73mm}
\begin{center}
\ig[height=0.3\tht]{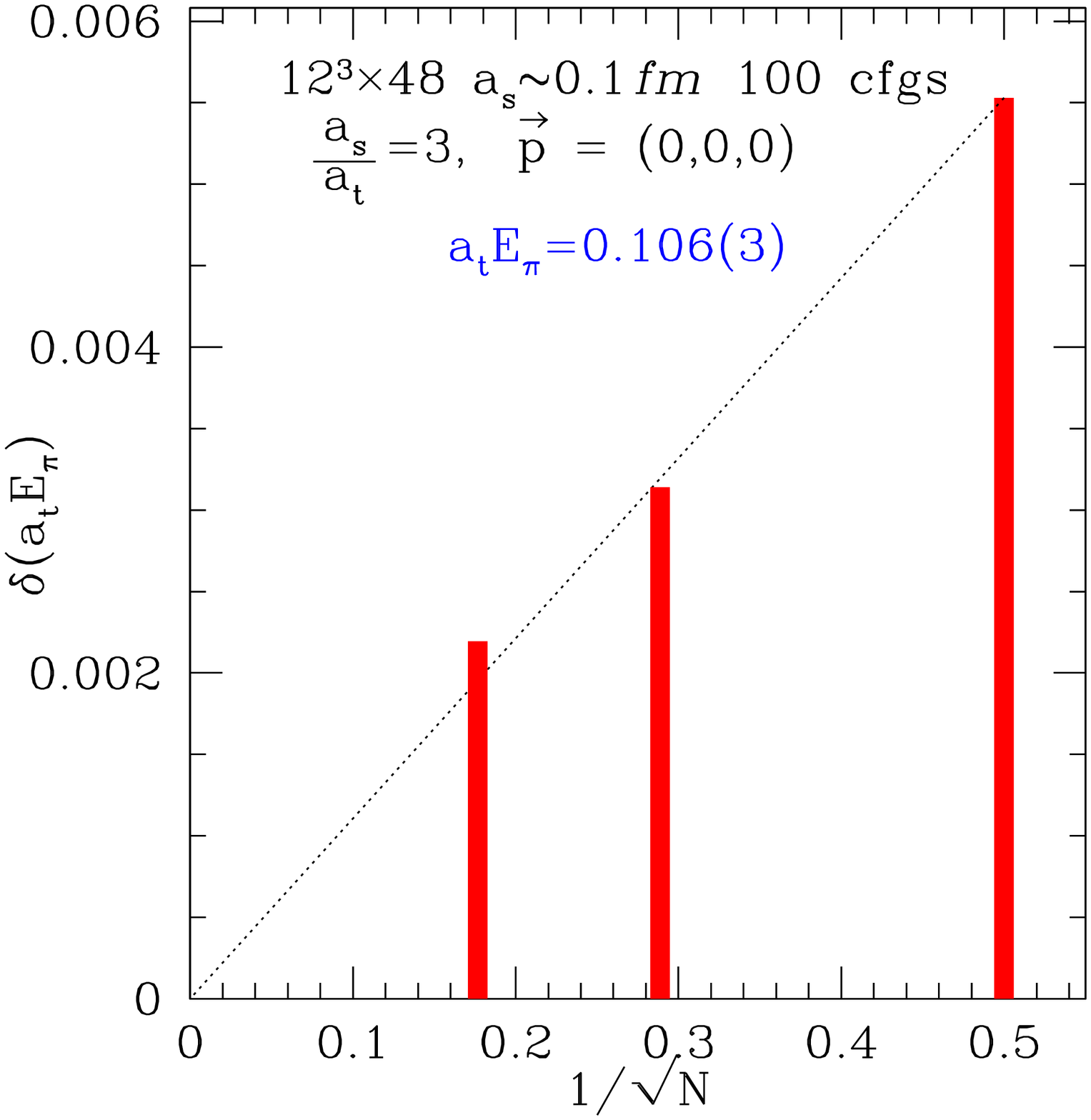} 
\caption{The error of the pion mass as a function of $1/\sqrt{N}$ where $N$ is the number of quark inversions needed in each dilution scheme.}
\label{fig:picomp}
\end{center}
\end{minipage}&
\begin{minipage}{73mm}
\begin{center}
\ig[height=0.3\tht]{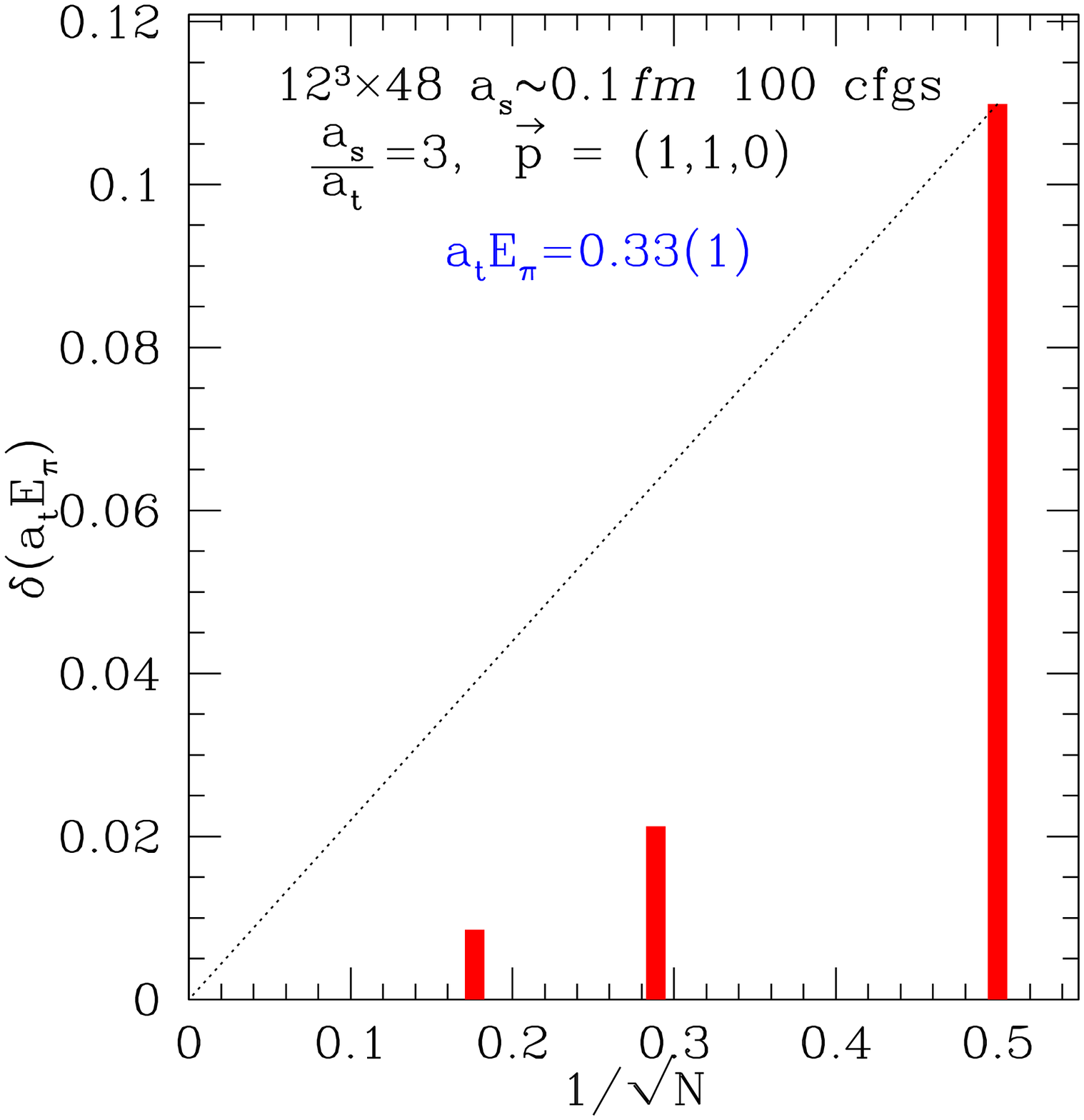}
\caption{The error of the pion energy with momenta $(1,1,0)$ (in lattice units) as a function of $1/\sqrt{N}$ where $N$ is the number of quark inversions needed in each dilution scheme.} 
\label{fig:pionfiniteP}
\end{center}
\end{minipage}
\end{tabular}
\end{center}
\end{figure}

The dilution scheme dependence of the error of the pion mass is shown in Fig.~\ref{fig:picomp}. The error is determined from a fit to the pion correlation function starting at $t_{min}=10$. This value of $t_{min}$ was not necessarily the optimal choice for all of the dilution schemes (in terms of $\chi^2$ etc) but a common choice is shown here for a lucid comparison. The dashed line is the line $1/\sqrt{N}$, which is the expected behaviour for $N$ repeated measurements with $N$ different noise pairs. The figure indicates that there may be better dilution choices than the ones investigated here (time+spin, time+spin+colour, time+spin+space-cubic). 

A similar plot for pions with finite momentum ($\vec{p}=(1,1,0)$) is shown in Fig.~\ref{fig:pionfiniteP}. The effect of colour dilution and in particular, the spatial-cubic dilution is self-evident. This result may have been anticipated, since the projection onto this particular momenta is not contaminated by the random noise when spatial (cubic) dilution is used. 

\subsection{Four-Quark Operators}
The dilution scheme dependence of the error of the $\pi^+\pi^+$ correlation function is shown in Fig.~\ref{fig:D-C}. The error in this case was the average of the errors in the region of $t$ where plateau for various dilution schemes were observed/expected. (We have used $5\le t\le 19$ in this example.) One can see from Fig.~\ref{fig:D-C} that space-dilution appears to be an effective dilution for scattering states, just as it was for pions with finite momenta. The cost comparison to the point-to-all case here is also difficult as the four-quark contraction is nontrivial there, but it is a factor of 4 more as far as the inversions are concerned for the stochastic case since there are four independent noise sources. 

\section{Summary}
The dilution method for approximating all-to-all quark propagators has been explored for two-quark operators with finite momentum and the simplest four-quark state ($\pi^+\pi^+$). We were able to obtain a signal for the six lowest momenta states for the pion with 100 configurations and spin$+$space-cubic dilution. Using the same operators that were used to measure the pion correlation function (i.e. without further simulations), we have demonstrated that a clean signal for the simplest four-quark state can be obtained with spin$+$space-cubic dilution. 

Simulations for five-quark operators are underway as well as the dynamical simulations of hadronic operators involving up to five quarks. A new all-to-all method for quark propagators which do not rely on stochastic estimates is also being explored. 

\section*{Acknowledgements}
K.J.J. would like to thank Gunnar Bali for many discussions on all-to-all propagators. 
This work has been partially supported by National Science Foundation awards PHY-0704171.
These calculations were performed using the Chroma software suite \cite{Edwards:2004sx} on clusters at Jefferson Laboratory using time awarded under the SciDAC Initiative and clusters at University of the Pacific. 


\end{document}